# Optical Hoovering on Plasmonic Rinks

JOHN CANNING,[1,2*]

[1] *interdisciplinary Photonics Laboratories, Tech Lab, School of Electrical & Data Engineering, The University of Technology Sydney (UTS), NSW 2007 & 2019, Australia*
[2] *School of Chemistry, The University of Sydney, NSW 2006, Australia*
*Corresponding author: john.canning@uts.edu.au

**Excitation of surface waves on conducting materials provides a near resistance-free interface capable of a material glissade either by plasmon forces or optical beam tractors. Analogous to an ice hockey rink, as proof of principle plasmon assisted optical traction, or hoovering, of water drops on a gold surface is demonstrated. Variability in thresholds and movement is observed and can be explained by the presence of significant roughness, measured by SEM. The demonstration opens a path to directly integrate various optical and plasmonic glissade technologies. Ways of improving transport and potential applications spanning configurable microfluidics, antennas, diagnostics, sensing and active devices are discussed. © 2018 Optical Society of America**

http://dx.doi.org/10.1364/OL.99.099999

Conducting surfaces with minimal binding energy along the interface, such that energy is easily coupled over a distance in the direction of an electronically or optically induced voltage, can behave analogously to that of a tightly bound ocean plasma. This is constrained to an *x,y* plane in which appropriate energy impartation, coupled between electrons, can create equivalent mechanical dynamics associated with localized surfaces waves shaped in part by the ocean landscape and a host of related phenomena. The generation of these surface wave resonances using phase matched optical coupling is achieved by a number of means. These include phase matching through prisms [1-3], end coupling using various optical wave guiding configurations and higher order modes or skew rays [4-6] and diffractive grating configurations [7-9]. The optical generation of surface waves, which are sensitive to their environment, has proven to be particularly useful for diagnostic applications [10,11]. Yet the potential for utilizing such excitation pathways can go well beyond exciting sensitive interactions. Here, the optical driven displacement of mass along such a surface is proposed and demonstrated.

On homogenous interfaces where, for example, resistive flow can be negated, physical displacement of an object on a surface should be possible. By exploiting a bound body of electrons capable of freely transferring kinetic energy in the direction of flow, a resistance-free surface may be generated and, with optical assistance, plasmon enabled movement of mass demonstrated – matter interacting with this interfacial sea should be capable of experiencing physical displacement, either plasmon, optical or a combination thereof.

In this work, metallic surfaces are considered, focusing on sputtered gold (Au) given it is widely used for surface plasmon resonance (SPR) generation and diagnostics. Au otherwise may not be the most suitable material because it can be difficult to make homogenously smooth surfaces at the scale required for reduced resistance at interfaces. Rather, other materials, such as 2D structures like graphene and silicene or even conducting liquids where roughness can be more readily reduced or eliminated, may be better suited. Nonetheless, an optical demonstration on Au where resistance may be high and the energy required greater, would be more challenging, further demonstrating what is possible.

Rather than solid microparticles, a water drop is used. As a liquid there is good contact, through both electrostatic and hydrogen bonding, leading to the likely presence of a thin ordered layer of water that stays on the surface long after most of the water drop appears to have evaporated [12]. The nature of these layers is a subject of significant intrigue – measurements on metal electrodes indicate self-assembled pentagon and hexagon-like arrays [13] whilst theoretical calculations suggest the layer at the interface has the array significantly broken in places with OH pointing towards the gold surface [15]. This latter interfacial contribution at the edges, along with surface roughness, can help account for the tendency of water to "pin" to a surface. This generates some resistance to ideal unconstrained flow, or glissade, on what might be a perfectly flat and hydrophilic surface. It may also explain the reduced electro-optic coefficients measured at metal-water interfaces compared to air-water interfaces on drops [12]. Water has also been used to highlight important microfluidic applications where what would ordinarily be the host medium supporting microparticles in conventional optical trapping work [16], is instead what is manipulated.

The components of the dielectric constant associated with a metal, for example, relate the physically distinct real parts of the refractive index (negative with respect to the dielectric) and the imaginary parts which, in metals, can involve energy loss through displacement and coupling (current) in the electron plasma. The complex dielectric is described as a sum of the two parts:

$$\tilde{\epsilon} = \epsilon' + i\epsilon'' = (n + i\kappa)^2 \quad \quad (1)$$

Light of *p*-polarisation impinging on a metal from the transparent substrate side will generate a time dependent polarization charge at the interface whereas *s*-polarised light does not. The plasma (ideally a non-bound charge layer in *x,y* space) does not have an instantaneous response resisting elastic dissipation of the coupled energy as a result of its association with the rigid surface and environment geometry, but also through inefficient shared energy transfer to other electrons. Even on an ideal perfectly flat surface, electrons have a finite mass and scatter energy as phonons or through defect states, including electron hole pairs, producing the resistive scattering measured as Ohmic loss (or Joule heating).

The direction of scatter is determined by the *k* vector of the impinging light and therefore controllable. This gives rise to a direct correlation between the SPR frequency and resistivity in the direction of flow. A corresponding "plasmo-motive" force might be anticipated.

The SPR measurements analysed here were carried out on a standard home-made Kretschmann setup (Figure 2) that has also been described elsewhere [3, 12, 17]. Surface roughness, as evidenced in Figure 1, broadens and weakens the SPR and greatly exacerbates this resistance due to complicated out-of-plane interactions (analogous to the rough geoformations under an ocean that impact undersea currents). In fact, such out-of-plane interactions alone can lead to additional, emergent angular excitation of SPR conditions [17]. Further, the two-dimensional nature of the surface also constricts perfect oscillations. Consequently, there are discrete or diffuse coupling conditions for many surface phenomena that need to consider these boundaries and practicalities. These may, for example, be guided within strip line configurations for desired device applications – they can be used to reduce the size of traditional antennas, including large bulky water antennas [18], for next generation 5G and beyond applications. Optimal energies below the plasma frequency exist where electrons can collectively respond, and dispersion is sufficiently high.

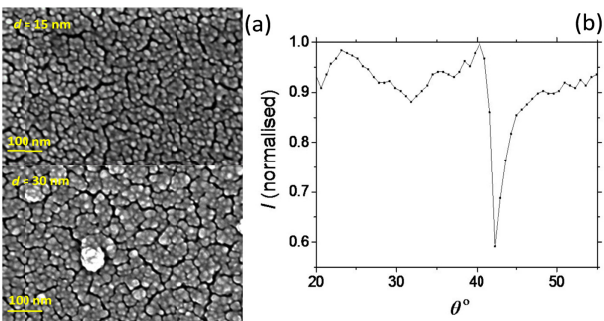

Fig. 1. (a) SEM images of two sputtered Au surfaces on borosilicate glass (15 nm and 30 nm); (b) Intensity spectra of reflected light from as sputtered Au surface on borosilicate ($\tau \sim$ 30 nm) measured in a Kretschmann SPR setup [17]. The broad SPR band and non-zero transmission dip at $\theta \sim 42°$ along with the background variation is an indication of significant surface roughness, consistent with SEM images.

Generally, the sensitivity to electron-hole interactions and geometric constraints makes the surface interface sensitive to environmental perturbations. Consequently, SPR based detection, a primarily fixed-point approach to date, is widely deployed. In many respects, however, a potentially closer analogy arises from the resemblance to the liquid interfacial state associated with water induced by physically imposed heating on a surface such as ice, one that permits movement of objects with reduced drag arising from the induced water layer. If the impartation on the layer of ice has a direction, then the body creating that impartation moves along accordingly with little resistance. In this analogy, it stands to reason the physical resistance of an object not chemically bound to the substrate under excitation should be greatly reduced and physical displacement with little energy possible. Such realization of physical movement enables an active approach to device fabrication and tuning of electronic, electromagnetic and physical properties.

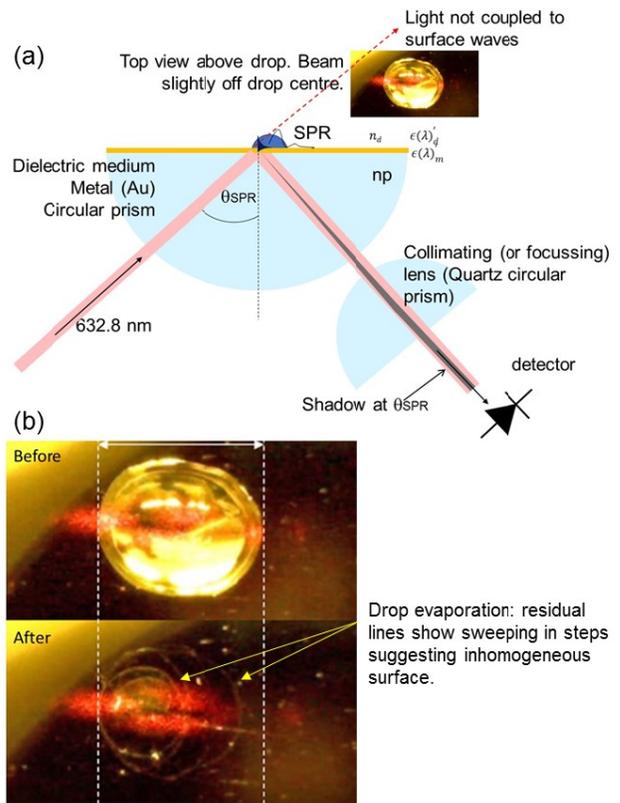

Fig. 2. Configuration used to excite surface plasmon resonances on gold films: (a) Kretschmann setup used in the experiments and (b) the observable transmission of light through $\tau \sim$ 15 nm layers. The HeNe laser used in this work has $P$ = 1 mW. That light matching the SPR condition is absorbed and is seen as a dark region in the centre after evaporation. Residual interactions between water and Au can also be seen: outer and an inner ring regions are revealed through illumination of the edges of interfacial water that resists evaporation. These are forensic evidence of movement towards the optical peak on the left.

In principle given the momentum coupling between light and the surface wave at Brewster-like conditions, the first experiments explored whether movement was feasible with this condition

alone – essentially a plasmonic tractor or tweezer effect. Whilst in the approach proposed this should be possible, it has not been observed to date in part because it has not been sought. It may also be considered very unlikely given the quality of many surfaces such as sputtered gold and the importance of minimal and homogenous frictional resistivity. To explore this, a relatively simple system was analysed. A drop of distilled water with a sufficiently small volume to fall within the surface-tension driven spherical caplet limit ($V \sim 2$ μL) [16] was placed onto a gold surface of the prisms used in the Kretschmann setup (Figure 2) optimized for SPR coupling with thicknesses of $\tau \sim 15$, 30 and 45 nm. In this limit, standard wide-view contact angle measurements are made in combination with SPR excitation allowing a simple but powerful surface characterisation technique. In practice, a miniature portable handheld microscope system was used to image the drops from the side when sitting on the Kretschmann setup. The drops had to be placed closer towards the prism edge to ensure good imaging. This combined method permitted a simple way to both monitor visual changes in drop shape and size along with changes in contact angle (manually measured on screen using microscope camera imaging software), arising from both evaporation and optical hoovering as described earlier.

For a review of the potency of contact angle, $\theta_c$, measurements see references [3, 12, 17, 19, 20]. Generally, the contact angle reflects the triple point interface between solid surface, liquid and air which pins the drop to the surface. The interface is extremely sensitive to chemical and physical properties from molecular to nanoscale and sometimes micro scale length interactions depending on surface viscosity of a liquid. They are a standard tool to characterise nanoscale surface interactions between liquids and solids, offering a measure of wettability (or attractiveness) between the two states of matter. For the sputtered Au used in this work, water tends to be hydrophilic where the contact angle is $\theta_c \sim 80°$ for pristine clean surface, typically dropping to $\theta_c \sim 70°$ or more after initial drop evaporation because a self-assembled layer on the surface can be more tightly bound [12]. A more recent variation extracting the contact angle from diameter measurements from the top rather than directly measuring from the side allows complete surface mapping of an essentially nanoscale property in real time [19,20]. It can also determine asymmetry in surface wettability, making it a particularly powerful technique for studying surface properties and behaviors. For example, it was more recently applied to mapping wettability in orchid leaves [20].

For initial SPR measurements, the 45 nm layer had no significant SPR profile consistent with the sample being too thick with little light reaching the surface. By displacing off-centre the relative position of the HeNe optical coupling below the drop of water, no corresponding displacement of the drop was observed from deposition through to full evaporation. Similarly, when light is transmitted at $\theta \sim 90^0$ for all layer thicknesses ($\tau \sim 15$, 30 and 45 nm), no displacement from optical traction alone is observed. These results are as expected and support the notion that the excitation of the SPR is important. Ideally, the same plasmo-motive force of the SPR leads to movement of a surface sample – unfortunately, at the close to optimal thickness ($\tau \sim 30$ nm) for maximum optical absorption and coupling of an input HeNe laser beam no visible displacement is observed, suggesting such a force is too small to overcome residual friction arising from surface variability.

The question then is if inhomogeneity and variability in surface roughness plays a key role in increasing resistivity across a surface, preventing displacement, could sufficiently greater force be utilized to overcome this resistance? In this situation, optical forces related to the interaction between a light beam and a dielectric or conducting dielectric (such as a self-assembled water layer) may be exploited. Ordinarily these optical forces are not considered substantive to move anything other than attract small micron-scale particles in unconstrained free space or within a flowing liquid. Given the potential of a plasmon-like sea to provide near frictionless movement of water, despite some resistance from surface variability, it should be feasible to physically displace well-constrained water along a metallic or conducting surface with additional force whilst simultaneously exciting the plasmon below. In this case, if it sits on an optically coupled surface wave whilst at the same time interacts with an excess of optical beam passing through the interface, then the attractive force of the beam on the dielectric may pull the drop of water towards it after a certain threshold is reached. This is the starting point of the proposal, further complicated in practice by both the mass of the water and its surface tension.

The Au layer is reduced in thickness ($\tau \sim 15$ nm) such that some HeNe light interacts with the metal to generate a surface plasmon whilst the rest passes through. Here, the standard side measurement technique is able to capture and demonstrate "optical hoovering", where the optical beam passing through sweeps up some of the water towards itself once a resistance to flow is overcome. This is effectively the beginnings of an optical tractor or tweezer on a chip coupled with an analogous plasmon sweep.

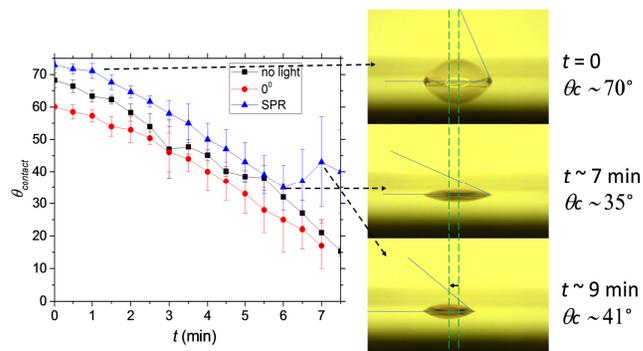

Fig 3. Progression of measured contact angle, $\theta_c$, with time, t, of a drop of water on Au film with thickness t $\sim$ 15 nm (graph on left). When water evaporated, a residual layer on the surface is left behind. It alters the starting contact angle by $\Delta\theta \sim 10^0$ (for pristine Au $\theta_c \sim 60^0$). After a threshold angle corresponding to surface resistance, an optical HeNe beam ($\lambda = 632.5$ nm) exciting surface plasmons slightly off to the left of the drop accumulates water towards the light that passes through – an enhanced optical tweezer effect enabled by creating a sliding electron layer at the Au/water interface. Such behavior is not observed away from the SPR condition and the magnitude varies with surface roughness.

When the Au layer is only $\tau \sim 15$ nm thick, significant transmission of HeNe light is observed except where the exact matching condition is met (Figure 2 b illustrates the off-centre excitation). In contrast with the thicker samples where little or no light is observed at all, only at the incident angle closest to SPR matching is there an attenuation within the beam profile where light can be seen not passing through. For this situation during evaporation, there was a distinct jump in water diameter observed after a finite period of evaporation. These jumps are directly observable as trace residue lines where varying pinning has occurred at the surface. The origin of the residue is unclear but arguably related to OH interactions with the surface being strongest at the drop edge. Despite Au's reputation as an inert material, deionized water can deprotonate on a Au surface, a property that is believed to explain how Au can catalyse reactions [21] and which leads to self-assembly. Since de-ionised water has been used, the residual traces of water can be related to strong deprotonation and electrostatic binding of water to the inhomogeneous surface. As well, interfacial van der Waals attachment of water to silica surfaces leads to nanoscale layers of strongly bound residual water that requires significant temperature (up to 40 $^0$C) to remove [12].

A summary of one set of results is shown in Fig 3. It shows clearly a reversing contact angle as water is pulled towards the optical beam. The jump in angle reflects the point at which resistance to movement is overcome during evaporation – this resistance should give information on surface inhomogeneity. It correlates with the jumps observed in Figure 2. At $\theta_c \sim 35°$ this occurs well before normal water pinning that prevents drop contraction during evaporation, is overcome ($\theta_c \sim 4$-$5°$). The rising contact angle with further evaporation indicates an optical pulling force more than strong enough to overcome surface pinning and hydrophilic spreading. A measure of how significant this effect, the net mass of a 2 µL drop far exceeds the typical particle sizes used in optical tweezers. For example, the trapping of polystyrene particles up to 130 µm (weight ~ 10 ng) within a flowing capillary required watts of power from a 5 W diode bar [22]. By comparison, in this work 2 µL of water (weight ~ 2 mg) has been swept with sub mW of optical power, a dramatic demonstration of the advantages of a frictionless support coupled with a contribution from a plasmon pushing force.

Experimentally, the amount of hoovering and timing during evaporation varied with different samples, some having lower resistance whilst others much higher. This was attributed to the variable induced surface roughness on the sputtered films which generally impacts the quality of the SPR. Its clear tremendous improvements are feasible with improvements in fabrication and optimisation of conditions. The latter can be aided by recent developments of top-down contact angle mapping that offer a way to scaling the number of measurements rapidly on a surface, enabling statistical information that can be used to both identify and reduce the source of errors [19,20].

In conclusion, this work has demonstrated the beginnings of an integrated form of optical tractor (or tweezer), taking advantage of surface wave resonances to reduce friction and allow movement of significant mass along a surface. A novel characterisation based on combining SPR generation with contact angle monitoring of water on Au was used. It has identified the potential of plasmo-motive forces for manipulating mass on a surface. The demonstration manipulates a significant body of water with little relative optical excitation power indicating that the entire system could be integrated into compact form on a chip, perhaps using multimode waveguide and waveguide device approaches [6,23] integrating a pump waveguide alongside a microfluidic channel. This approach has immediate use in dynamic diagnoses, combining potentially tuneable lateral flow and novel separation with various spectroscopies or probes. Mindful of the extraordinary potential of water as a novel platform material for electro-optics [24], the potential to manipulate, move and tune water properties offers exciting new possibilities in lensing, switching and more. The methods described here can also be applied to other liquids or solids that can be glided along the surface.

**Funding.** Australian Research Council (ARC) (DP 140100975); Private Funding.

**Acknowledgment**. Ideas, project direction, analysis and understanding are the Author's. Special thanks go to Mr. Chunyang Han, a visiting China Scholarship Council (CSC) student at the University of Sydney, for carrying out the SPR experimental work reported in this paper and to Dr. Kevin Cook who took the SEM images reproduced here and helped organise the gold sputtering for Mr. Han. Both SEM and sputtering facilities were made available at cost from the Microanalysis Centre of the University of Sydney.

**References**

1. Otto, Z. Phys. **216**, 398 (1968).
2. E. Kretschmann, Z. Phys. **241**, 313 (1971).
3. J. Canning, J. Qian, and K. Cook, Phot. Sensors **5**, 278 (2015).
4. H. J. M. Kreuwel, P. V. Lambeck, J. V. Gent, and Th. J. A. Popma, Proc. SPIE **V798**, 218 (1987).
5. R. C. Jorgenson, and S.S. Yee, Sensors and Act. B Chemical, **12**, 213 (1993).
6. C. Han, J. Canning, K. Cook, M.A. Hossain, and H. Ding, Opt. Lett. **41**, 5353 (2016).
7. P. S. Vukusic, G. P. Bryan-Brown, and J. R. Sambles, Sensors and Act. B: Chemical **8**, 8155 (1991).
8. J. Dostalek, J. Homola, and M. Miler, Sens. Actuators B **107**, 154 (2005).
9. R.W. Wood, Philos. Mag. **4**, 396 (1902).
10. J. Homola, S.S. Yee and G. Gauglitz, Sensors and Act. B: Chemical, **54**, 3 (1999).
11. Z. Altintas, and W. M. Fakanya in *Biosensors and Nanotechnology: Applications in Health Care Diagnostics* (Ed. Z. Atlintas 2017), Ch 4.
12. J. Canning, N. Tzoumis, J. Beattie, Chem. Commun., **50**, 9172 (2014).
13. E. Tokunaga, Y. Nosaka, M. Hirabayashi and T. Kobayashi, Surface Science **601**, 735 (2007).
14. Y. Suzuki, K. Osawa, S. Yukita, T. Kobayashi, and E. Tokunaga Appl. Phys. Lett. **108**, 191103 (2016).
15. M. J. Stevens and G. S. Grest, Biointer*phases* **3**, FC13 (2008).
16. C. Bradac, Adv. Opt. Materials **6**, 1800005 (2018).
17. J. Canning, A. Karim, N. Tzoumis, Y. Tan, R. Patyk, and B. C. Gibson, Opt. Lett. **39**, 5038 (2014).
18. H. Fayad, P. Record, Electron. Lett. **42**, 133 (2006).
19. G. Dutra, J. Canning, W. Padden, C. Martelli, and S. Dligatch, Opt. Express **25**, 21127 (2017)
20. C. Janeczko, C. Martelli, J. Canning and G. Dutra, "Assessment of orchid surfaces using top-down contact angle mapping", Accepted to IEEE Access, (2019).
21. J. Saavedra, H. A. Doan, C. J. Pursell, L. C. Grabow, and B. D. Chandler, Science, **345**, 1599, (2014).
22. R. W. Applegate Jr., D. W. M. Marr, J. Squier, and S. W. Graves, Opt. Express, **17**, 16731 (2009).
23. M. Åslund, S. D. Jackson, J. Canning, A. Texeira and K. Lyytikäinen, Optics Comm., **262** (1), 77-81, (2006).
24. J. Canning, MRS Commun. **8**, 29 (2018).

1. A. Otto, "Excitation of non-radiative surface plasma waves in silver by the method of frustrated total reflection", Z. Phys., 216, 398-410 (1968).
2. E. Kretschmann, "The determination of the optical constants of metals by excitation of surface plasmons," Z. Phys. 241, 313-324 (1971).
3. J. Canning, J. Qian, K. Cook, "Large dynamic range SPR measurements in the visible using a ZnSe prism", Phot. Sensors, **5** (3), 278-283, (2015).
4. H.J.M. Kreuwel, P.V. Lambeck, J.V. Gent, and Th.J.A. Popma, "Surface plasmon dispersion and luminescence quenching applied to planar waveguide sensors for the measurement of chemical concentrations", Proc. SPIE, V798, 218-224, (1987).
5. R.C. Jorgenson, S.S. Yee, "A fiber-optic chemical sensor based on surface plasmon resonance", Sensors and Act. B, 12, (3), 213-220, (1993).
6. C. Han, J. Canning, K. Cook, M.A. Hossain, H. Ding, "Exciting Surface Waves on Metal-Coated Multimode Optical Waveguides using Skew Rays", Opt. Lett., **41** (22), 5353-5356, (2016).
7. P.S. Vukusic, G.P. Bryan-Brown and J.R. Sambles, "Surface plasmon resonance on gratings as a novel means for gas sensing", Sensors and Act. B, **8** (2), 8155-160, (1991).
8. J. Dostalek, J. Homola, M. Miler, "Rich information format surface plasmon resonance biosensor based on array of diffraction gratings", Sens. Actuators B 107, 154-161 (2005).
9. R.W. Wood, "On a remarkable case of uneven distribution of light in a diffraction grating spectrum", Philos. Mag. 4, 396-402 (1902).
10. J. Homola, S.S. Yee and G. Gauglitz, "Surface plasmon resonance sensors: review", Sensors and Act. B, 54, 3-15, (1999).
11. Z. Altintas, W. M. Fakanya, "SPR-Based Biosensor Technologies in Disease Detection and Diagnostics", Biosensors and Nanotechnology: Applications in Health Care Diagnostics (Ed. Z. Atlintas), Ch 4, (2017)
12. J. Canning, N. Tzoumis, J. Beattie, "Water on Au sputtered films", Chem. Commun., 50, 9172-9175, (2014).
13. E. Tokunaga, Y. Nosaka, M. Hirabayashi and T. Kobayashi, "Pockels effect of water in the electric double layer at the interface between water and transparent electrode", Surface Science **601**, 735–741 (2007).
14. Y. Suzuki, K. Osawa, S. Yukita, T. Kobayashi, and E. Tokunaga, "Anomalously large electro-optic Pockels effect at the air-water interface with an electric field applied parallel to the interface", Appl. Phys. Lett. **108**, 191103 (2016).
15. M. J. Stevens and G. S. Grest, "Simulations of water at the interface with hydrophilic self-assembled monolayers", Biointer*phases* 3 (3), FC13 -22, (2008)
16. C. Bradac, "Nanoscale optical trapping: A review", Advanced Optical Materials, 6 (12), 1800005 (2018)
17. J. Canning, A. Karim, N. Tzoumis, Y. Tan, R. Patyk, and B.C. Gibson, "Near orthogonal launch of SPR modes in Au films," Opt. Lett. 39, 5038-5041 (2014).
18. H. Fayad, P. Record, "Broadband liquid antenna", *Electron. Lett.* **42** (3), 133 (2006)
19. G. Dutra, J. Canning, W. Padden, C. Martelli, and S. Dligatch, "Large area optical mapping of surface contact angle," Opt. Express **25**, 21127-21144 (2017)
20. J. Saavedra, H. A. Doan, C. J. Pursell, L. C. Grabow, and B. D. Chandler, "The critical role of water at the gold-titania interface in catalytic CO oxidation", Science, 345 (6204), 1599-1602 (2014)
21. C. Janeczko, C. Martelli, J. Canning, G. Dutra, "Assessment of orchid surfaces using top-down contact angle mapping", Accepted to IEEE Access, (2019).
22. R. W. Applegate Jr., D. W. M. Marr, J. Squier, and S. W. Graves, "Particle size limits when using optical trapping and deflection of particles for sorting using diode laser bars", Opt. Express, 17 (19), 16731-16738, (2009)
23. M. Åslund, S. D. Jackson, J. Canning, A. Texeira, K. Lyytikäinen, "The influence of skew rays on the angular losses of air clad fibres", Optics Comm., **262** (1), 77-81, (2006)
24. J. Canning, "Water photonics, non-linearity and anomalously large electro-optic coefficients in poled silica fibres", MRS Commun. **8** (1), 29-34, (2018).